\documentstyle[aps,prl,multicol,epsf]{revtex}
\def\beq{\begin{equation}}
\def\eeq{\end{equation}}
\begin{document}                
\title{Possible Glassiness in a Periodic Long-Range Josephson Array}
\author{P. Chandra}
\address{NEC Research Institute, 4 Independence Way, Princeton NJ 08540}
\author{L. B. Ioffe}
\address{Department of Physics, Rutgers University, Piscataway, NJ 08855 \\
and Landau Institute for Theoretical Physics, Moscow}
\author{D. Sherrington}
\address{Department of Physics, Oxford University, 1 Keble Rd, Oxford,
OX1 3NP, UK}
{}
\maketitle
\begin{abstract}

We present an analytic study of a periodic Josephson array
with long-range interactions in a transverse magnetic field.
We find that this system exhibits a first-order
transition into a phase characterized by an
extensive number of states separated by barriers that scale with
the system size;
the associated discontinuity is small in the limit of weak applied field,
thus permitting an explicit analysis in this regime.

\end{abstract}

\begin{multicols}{2}

The essential features underlying the physics of glass formation are
still unclear.
In particular the possibility of glassiness in the absence of disorder
is a question that has stimulated much recent activity\cite{Parisi}.
In this paper we present and  study a periodic
Josephson array with long-range interactions
frustrated by a transverse magnetic field ($H$).
We show that it displays a first-order transition into a
low-temperature
phase characterized by an extensive number of states;
these metastable solutions are separated by barriers that
scale with the system size.  The discontinuity at the
transition is small in the limit of weak applied field,
so that an explicit analytic treatment is possible in this
regime.

\begin{figure}
\centerline{\epsfxsize=5cm \epsfbox{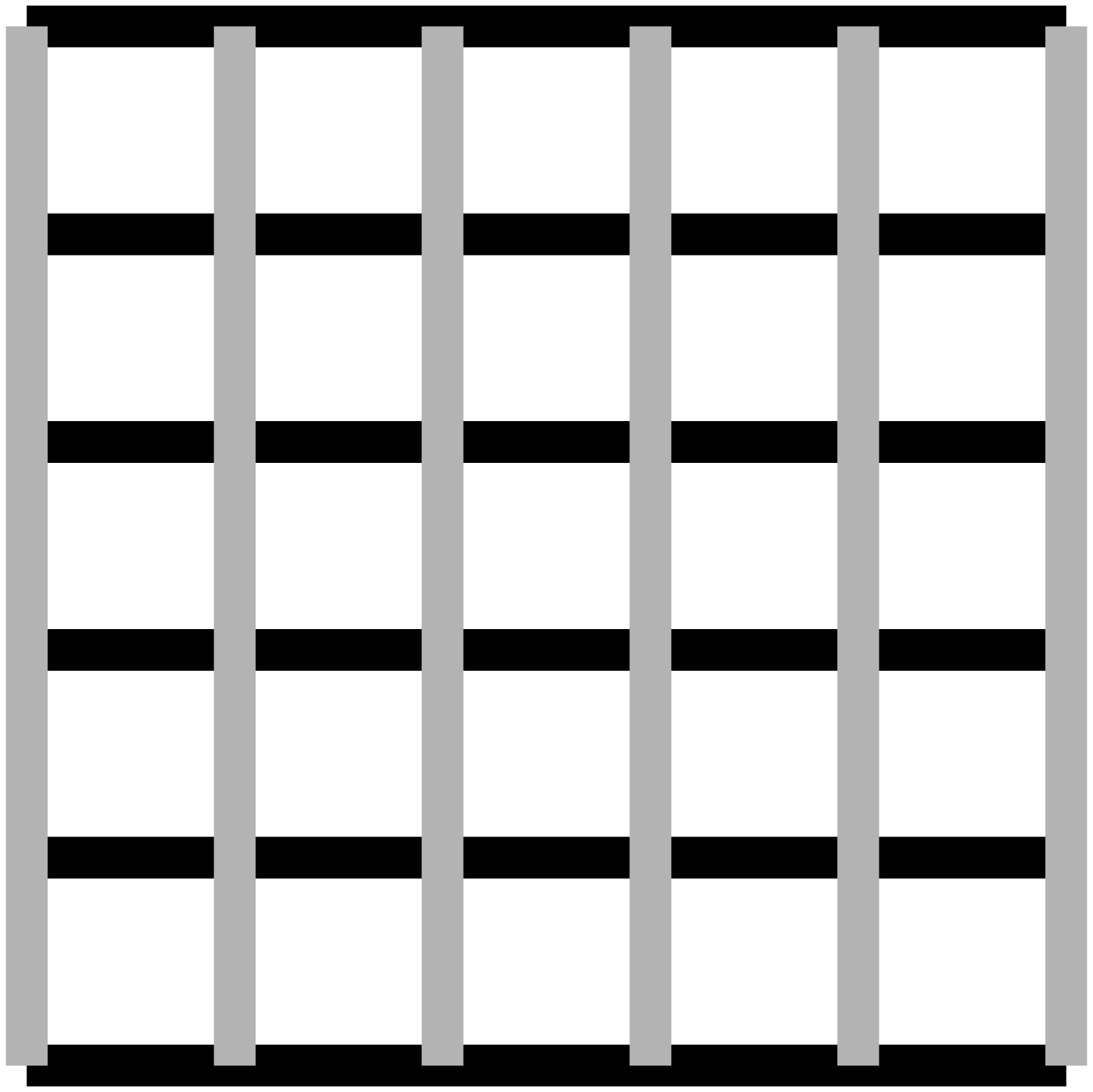}}
\end{figure}
Fig 1. A schematic of the array discussed in the text;
the solid (shaded) lines are
the horizontal (vertical) superconducting wires that are coupled by
Josephson junctions to the vertical (horizontal) wires at each node.

More specifically, the proposed  array is a stack
of two mutually perpendicular sets
of $N$ parallel wires with Josephson junctions at each node
(Figure 1)  that is placed in an external tranverse field.
The classical thermodynamic variables of this system are the
superconducting phases associated with each wire.
Here we shall assume that the Josephson couplings are sufficiently
small so that the induced fields are negligible in
comparison with $H$.
We can therefore describe the array by the Hamiltonian
\beq
{\cal H} =
- {\rm Re} \sum_{j,k}^N S_j J_{jk} S_k
= - \sum_{m,n}^{2N} s_m^{*} {\cal J}_{mn} s_n
\label{H}
\eeq
where ${\cal J}_{mn}$ is the coupling matrix
\beq
\hat{\cal  J} = \left( \begin{array}{cc}
0 & \hat{J} \\
\hat{J}^\dagger & 0
\end{array}
\right)
\label{J}
\eeq
with $J_{jk} = \frac{J_0}{\sqrt{N}} \exp(2\pi i \alpha jk /N)$
and
$S_j(s_m)$ is  a $N(2N)$-element vector whose components are $\exp(i\phi_m)$,
where $\phi_m$ are the superconducting phases of the wires.
Here we have introduced the flux per unit strip,
$\alpha = NHl^2/\phi_0$,
where $l$ is the inter-node spacing,
$\phi_0$ is the flux quantum,  and
the normalization has been chosen so that $T_c \sim J_0$.

Because every horizontal (vertical) wire is linked to every
vertical (horizontal) wire, the number of nearest neighbors in this
model $z = N$;
we can therefore study it with a mean-field approach.
For the same reason the free energy barriers
separating its low $T$ solutions scale with $N$, in marked
contrast to the situation in a conventional $2D$ array \cite{remark};
thus each of these solutions results in a metastable state.
If the field is such that the number of solutions is extensive
(i.e. the phase-ordering unit cell scales with the system size) this
array displays glassy behavior in the absence of disorder.

A similar long-range network with areal disorder
was studied previously\cite{disorder}; for $\alpha \gg 1/N$ the
system undergoes a spin glass transition
which was mapped onto the
Sherrington-Kirkpatrick model for the case
$\alpha \gg 1$ for all $T$.
Physically this behavior occurs because the phase differences
associated with the couplings, $J_{jk}$, acquire random values and
fill the interval $(0,2\pi)$ uniformly for $\alpha \gg 1/N$.
In the absence of randomness these phases still cover the interval for
$\alpha \gg 1/N$, so that one expects that glassy behavior survive
\cite{ferro}.
Naturally, there will be special values of the field for which the
number of solutions is finite.
However there is no such commensurability for $1/N \ll \alpha \leq 1$;
e.g. the unit cell corresponding to $\alpha = 1$ consists of all the
wires traversing a strip and thus has sides of order $O(N)l$.

The cumulants of the coupling constants display a more quantitative
similarity between the random and the periodic arrays.
In both cases
\beq
\sum_{j_1,i_2 ... j_n}
 J_{i_1 j_1} \ldots J_{j_n i_1}
=J_0^{2n}
	\left(\frac{1}{\alpha}\right)^{n-1}
\label{cumulants}
\eeq
which can be understood physically since only horozontal (vertical)
wires separated by $d < l/\alpha$ contribute coherently
to this sum independent of disorder.

We now begin a more quantitative analysis of this periodic array using
a modified Thouless-Anderson-Palmer approach \cite{TAP}.
Because the interactions are long-range, we can
integrate out the thermal fluctuations and define a ``site
magnetization'' $m_i = \langle z_i \rangle_T$.
The free energy is
\begin{eqnarray}
F\{ m_k \} &=& - \sum_{kj} m_k^* {\cal J}_{kj} m_j - T \sum_k S(m_k) +
	F_o\{m_k\}
\nonumber \\
	& & - \sum_k (h_k m_k^* + h_k^* m_k)
\label{F} \\
S(m) &=& S_0 - |m|^2 - \frac{1}{4} |m|^4 + O(|m|^6)
\nonumber
\end{eqnarray}
where ${\cal J}_{kj}$ is defined in (\ref{J}).
Here we have introduced $h_i$, a field conjugate to $m_i$, and the
single site entropy $S(m)$; $F_o\{m_k\}$ is the Onsager reaction term.
At high temperatures $F_o\{m_k\} \sim 1/T$ and thus is small;
furthermore, as we show below, the transition occurs at
$T_c \approx J_0/\sqrt{\alpha}$ (as in the disordered case) so
that we expect the feedback effects to be perturbative for small
$\alpha$ ($1/N \ll \alpha \ll 1$).
Thus we shall ignore $F_o\{m_k\}$ in our initial study of $F\{m_k\}$,
but will consider it later.

In the high temperature phase $m_i \equiv 0$.
This ``paramagnetic'' state becomes unstable when the quadratic part
of $F\{m_k\}$ acquires negative eigenvalues, i.e. at $T \leq
\lambda_{max}$ where $\lambda_{max}$ is the largest eigenvalue of
$\hat{\cal J}$.
Due to the structure of $\hat{\cal J}$ (see (\ref{J})), any eigenvalue
$\Lambda$ of $\hat{J} \hat{J}^\dagger$ results in two eigenvalues $\lambda=\pm
\sqrt{\Lambda}$ of $\hat{\cal J}$, a property we exploit for
convenience. In the limit $N \rightarrow \infty$
\beq
\left( \hat{J} \hat{J}^\dagger \right)_{jk} =
	J_0^2 \mbox{e}^{i \pi \alpha(j-k)}
	\frac{\sin(\pi\alpha(j-k))}{\pi \alpha (j-k)}
\label{JJ}
\eeq
which can be diagonalized by a Fourier transformation
to yield
\beq
\left( \hat{J} \hat{J}^\dagger \right)_{p} = \frac{J_0^2}{\alpha}
	\theta(p) \theta(2\pi\alpha -p).
\label{JJ_p}
\eeq
Therefore the largest eigenvalue of $\hat{\cal J}$ is
$\lambda_{max}=J_0/\sqrt{\alpha}$; the corresponding eigenfunctions
associated with one set of parallel wires are
plane waves with momenta in the interval $0 \leq p \leq 2\pi \alpha$
so that the degeneracy of this eigenvalue is $\alpha N$\cite{numerics}.

Therefore, in the absence of the feedback effects, the transition
occurs at $T_{c0}=J_0/\sqrt{\alpha}$ into a phase that is characterized
by a linear combination of the $\alpha N$ states described above.
Thus the number of possible metastable configurations just below the
transition should be extensive, indicating the onset of glassiness
\cite{disclaimer}.
This should be contrasted with the situation in the
Sherrington-Kirkpatrick model where, in the thermodynamic limit, the
highest eigenvalue of the exchange matrix is non-degenerate and the
metastability measure of the order parameter (plateau-onset $x_0$ in Parisi
function $q(x)$) grows continuously beneath the transition.
More generally, in disordered spin glasses a sudden appearance of
extensive metastability implies a first-order jump in the order
parameter \cite{Sherrington}.
Unfortunately in the periodic case we have not yet determined the
analogous order parameter; however we show below that the Onsager
feedback term drives the transition weakly first-order, consistent with
our expectations.

We use the locator expansion \cite{locator} to determine the leading
order contributions in $m_i$ to $F_o\{m_k\}$,
it is based on the expression for susceptibility
\beq
\hat{\chi} = \frac{1}{\hat{A} - \hat{\cal J}}
\label{chi}
\eeq
where $\hat{A}$ is a diagonal ``locator'' matrix.
This form of susceptibility is a consequence of large $z$; to prove it
one can use the high temperature expansion and show that
all diagrams renormalizing the $\hat{\cal J}$ term are small in $1/z $.
Once we have solved for $\hat{A}$, we can reconstruct the free energy
from
\beq
\frac{\partial^2 F}{\partial m_k^{*} \partial m_j}
=\left(\hat{\chi}^{-1}\right)_{kj}
\label{F_eq}
\eeq
that gives us the free energy which contains both the entropy term $\sum
TS(m_i)$ and the Onsager term $F_o\{m_i\}$.
Although it is possible, in principle, to determine the elements of
$\hat{A}$ {\em ab initio} it is easier to do so using the identity
$\chi_{jj} \equiv \frac{1-|m_j|^2}{T}$ which follows from the
fluctuation-dissipation theorem.
We expand the expression for the susceptibility (\ref{chi}) and, due
to the structure of $\hat{\cal J}$, find that only terms even in
$\hat{\cal J}$ contribute.
Collecting and resumming these terms, we obtain
\beq
\left( \frac{1}{\hat{A} - \hat{J} \hat{A}^{-1} \hat{J}^\dagger}
	\right)_{jj} =\frac{1-|m_j|^2}{T}.
\label{A_matrix_eq}
\eeq
We first consider the leading-order term (setting $m_j=0$) which will
yield the quadratic part of $F\{m_j\}$.
To this order $\hat{A}$ is independent of the site label ($\hat{A} = A
\hat{1}$), so it is convenient to solve this equation by Fourier
transformation and, using (\ref{JJ}), we obtain
\beq
\frac{\alpha}{A-\frac{J_0^2}{A \alpha}} + \frac{1-\alpha}{A} =
\frac{1}{T}
\label{A_eq}
\eeq
The zeroth-order term in $\alpha$ yields $A_0=T$, consistent with our
assumption above that the reaction term is small in $\alpha$.

\begin{figure}
\centerline{\epsfxsize=7cm \epsfbox{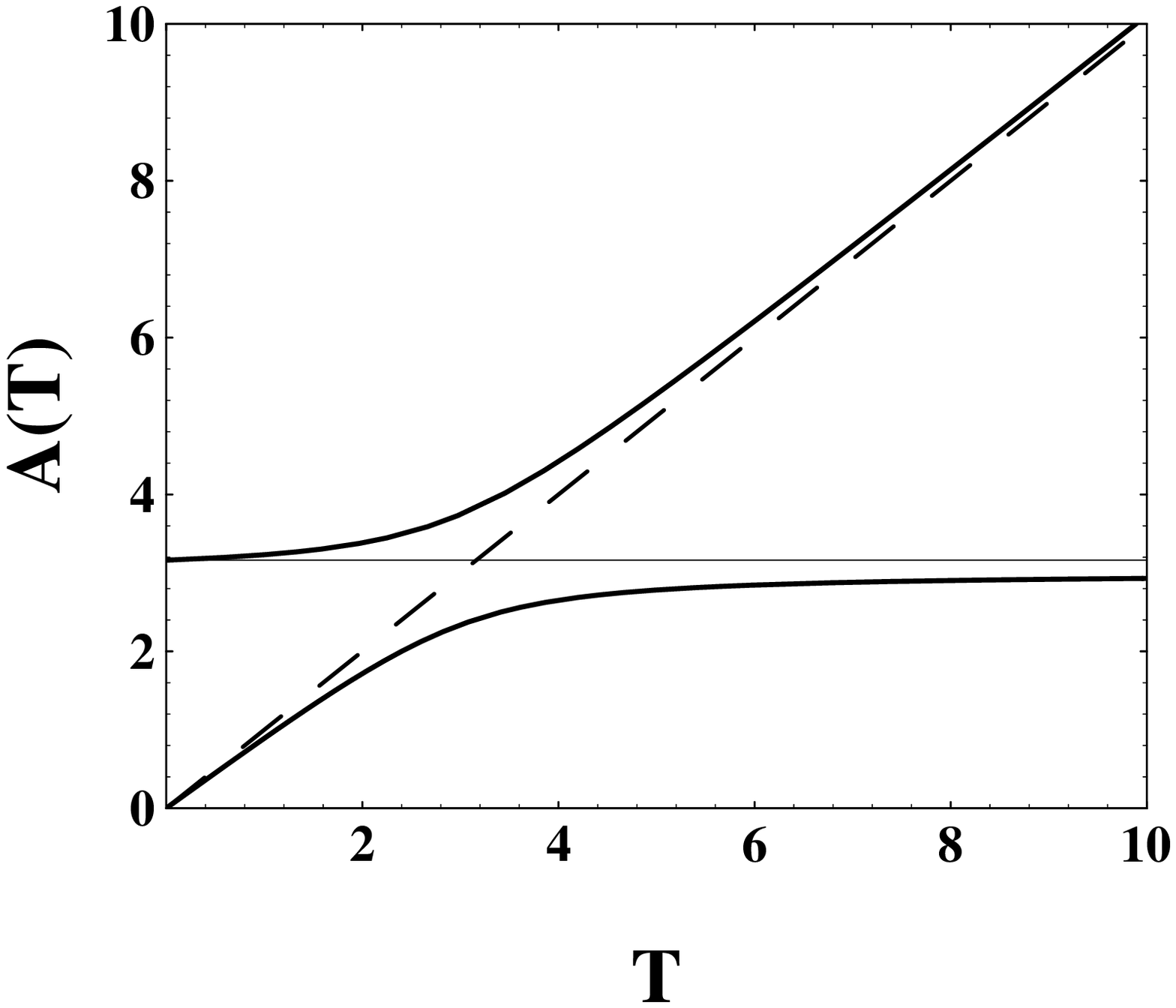}}
\end{figure}
Fig 2. Plot of the locator $A(T)$ obtained as a numerical solution of
Eq. (10) for $J_0=1$, $\alpha=0.1$; the horizontal line denotes the
critical value  $A(T)=\lambda_{max}=J_0/\sqrt{\alpha}$.
\vspace{0.2in}

The solution  $A(T)$ has two branches $A^{\pm}(T)$ displayed in Figure 2.
Since $A(T)$ is always larger than $J_0/\sqrt{\alpha}$ on the upper
branch ($A^{+}(T)$), within this approximation the high temperature phase is
{\em locally} stable down to zero temperature.
It is therefore impossible to reach the low temperature phase from
high temperatures without a discontinuity in $A(T)$.
A lower limit on the discontinuity is the gap between two branches $\delta A =
\frac{1}{2} \sqrt{\alpha} J_0 \ll J_0/\sqrt{\alpha}$ (recall that we
are taking $\alpha \ll 1$).
We note that although the feedback terms are generally small, they
become important near the transition and here change its order.
We can estimate the corresponding {\em minimal} jump in $|m_i|^2$
neglecting the feedback effects
on the quartic terms (i.e. using $F^{(2)} =[A(T)-\lambda_{max}] |m|^2$,
$F^{(4)} = \frac{1}{4} T |m_i|^4$),
to be $\delta |m_i|^2 = \alpha$.
In the vicinity of $T_{c0}=\lambda_{max}$ the discontinuity becomes
larger ($\Delta A=\lambda_{max}-A^{-}(T_{c0}) = J_0/\sqrt{2}$); since we
expect the
transition to occur in the vicinity of $T_{c0}$, the most likely value of
magnetization jump is $\Delta |m_i|^2 = \sqrt{2\alpha}$, as is also borne
out by a more sophisticated calculation \cite{details}.

In the regime of strong frustration, when $\alpha \sim 1$, the
transition becomes strongly first order.
The case $\alpha=1$ has a special interest because in this situation
$\hat{J}$ becomes a unitary matrix and cumulants of $\hat{\cal J}$
resemble those of the SK model: $\frac{1}{2N} \mbox{Tr} \hat{\cal
J}^{2n} = J_0^{2n}$.
As a result, the calculation becomes simpler.
The leading order locator equation (\ref{A_eq}) can be solved
explicitly
\beq
A^{\pm}_{\alpha=1} = \frac{1}{2} [T \pm \sqrt{T^2+4J_0^2}]
\eeq
suggestive of a strong first order transition around $T \sim J_0$.
Certainly, the existence of such a transition cannot be determined by a
$m$-expansion.
Instead, we adopt a variational approach and demonstrate that there exists
a ``magnetic'' phase  whose free energy is lower
than the ``paramagnetic'' free energy $F^{+}=-2N T S_0(T)$ associated with
the upper branch $A^{+}(T)$.
We choose the trial state
\begin{eqnarray}
m^{(v)}_j &=& \exp(-i \frac{\pi}{N} \gamma j^2)
\label{mv_trial} \\
m^{(h)}_k &=& (i)^{1/2} \exp(i \frac{\pi}{N} \gamma^{-1} k^2)
\label{mh_trial}
\end{eqnarray}
where $\gamma$ is a variational parameter;
since $|m_j|^2=1$ the free energy of this state is equal to its
energy
\begin{eqnarray}
E &=& - J_0 \sum_{j,k} \frac{(-i)^{1/2}}{\sqrt{N}} \exp[-i\frac{\pi}{N}
(\gamma j - k/\gamma)^2] + c.c.
\nonumber \\
  &=& -2N \epsilon(\gamma) J_0
\label{E}
\end{eqnarray}
where the dimensionless function $\epsilon(\gamma)$ satisfies
$\epsilon(\gamma) = \epsilon(1/\gamma)$
and $\epsilon(\gamma) = \gamma$ at $\gamma \ll 1$.
Clearly at $T\ll J_0$ this state has a lower free energy than the
paramagnetic state associated with $A^{+}(T)$, and thus we conclude that
it is unstable at low temperatures.

We emphasize that it is difficult to observe a paramagnetic
instability in  the free energy  expanded in $m_i$; we have already seen that
its quadratic part remains positive definite on the
upper branch $A^{+}(T)$.
Now we show that this conclusion does not change when quartic terms
are taken into account.
To do so  we assume the most general form of the
free energy possible at $\alpha=1$
\begin{eqnarray}
F\{m_j\} &=& - \! \sum_{kj} m_k^{*} {\cal J}_{kj} m_j + \sum_j [ A(T)|m_j|^2
+ \frac{1}{4} b |m_j|^4 ]
\nonumber \\
         &-& \! \sum_j (h_j^{*} m_j + c.c.)
+ \frac{1}{2N} \sum_{kj} a_{kj} |m_k|^2 |m_j|^2
\label{F4}
\end{eqnarray}
where the coefficients $a_{kj}=a_{+}$ if $\{k,j\}$  label parallel
 wires and $a_{kj}=a_{-}$ otherwise.
We
solve the equation $\frac{\partial F\{m_j\}}{\partial m_k} = 0$
to find the nonlinear susceptibility $\chi^{(3)}_{kj} \equiv
\frac{\partial^3 m_k}{\partial h_k \partial h_j \partial h_j^{*}}$;
it also satisfies
$\chi^{(3)}_{kj} = - \frac{1}{T} |\chi_{kj}|^2 + O(|m|^2)$.
Comparing these two expressions we  determine the coefficients $a_{+}$,
$a_{-}$ and $b$:
\beq \begin{array}{rcl}
b     &=& T \\
a_{+} &=&    \frac{J_0^4}{A^4(T)-J_0^4} T \\
a_{-} &=& -  \frac{J_0^2 A^2(T)}{A^4(T)-J_0^4} T.
\end{array}
\label{ab}
\eeq
The quartic term of the free energy is minimized when $|m_j|^2=m^2$, but
still  remains positive for the upper branch
($A^{+}(T)>J_0$)
\beq
F(m) = 2N \left[ (A(T)-J_0^2)m^2 + \frac{1}{4} T
\frac{A^2(T)-J_0^2}{A^2(T)+J_0^2}
	m^4 \right]
\label{F(m)}
\eeq
implying that all solutions remain stable to quartic order.

In summary, we have presented a non-random model which exhibits a
first-order transition to a phase where the number of metastable
states is extensive, consistent with one's general expectations for
glassiness in the absence of disorder\cite{Parisi}.
In this array the interaction between wires is long-range, and thus
the barriers separating the low-temperature metastable states scale
with the size of the system.
Since the number of nearest-neighbors $z$ is large we could
perform an initial investigation  using a modified TAP approach;
furthermore the glass transition is weakly first-order in the
limit of small $\alpha$
($\frac{1}{N} \ll \alpha \ll 1$), the flux per unit strip, so that
it could be studied analytically.

The results presented here can be tested in numerical simulations.
The system will have a signatory superfluid
stiffness, $\rho_s$,  which can be measured by applying a twist $\Phi$ to the
the horizontal(vertical) wires at their
respective boundaries\cite{Teitel}; we predict that $\rho_s =
\frac{\partial^2 F}{\partial \Phi^2}$  will appear discontinuously at the
transition $T_c(\alpha)$ and that the jump will scale with $\alpha$.
Thermal cycling should indicate an extensive number of metastable
states, thereby confirming the low-temperature glassy phase predicted here.

This transition could also be probed in fabricated arrays
with negligible induced fields; more specifically we require
\begin{equation}
J_0 < \frac{\Phi_0^2}{N^{5/2} l}
\label{J_0}
\end{equation}
which constrains the size of the array\cite{Tinkham}.
Here $T_c(\alpha)$ should be observable by resistive measurements.
Alternatively one can infer the superfluid stiffness from inductive
experiments.
The critical current below $T_c(\alpha)$ should jump as well.

Of course, a glass transition is best characterized by its dynamical
features, e.g. relaxation spectrum, hysteresis, ageing and memory effects;
we have
not yet investigated these properties of the periodic array discussed here.
Hopefully the order parameter associated with this glass transition
will also emerge from such a study,
since we expect it to be a dynamical quantity.
The thermodynamic behavior of a similar array with a reduced
number of neighbors is another open question which could
be important in understanding the relevance of long-range models
to experimental glasses.

We would like to acknowledge M.V. Feigel'man for several useful
discussions.  We also thank the
University of Oxford, the Aspen Center for Physics and NEC
Research Institute for their hospitality.
Partial financial support at Oxford
was provided by SERC UK under Grant GR/H6680.

\end{multicols}


\begin{references}
\bibitem{remark} We also note that in contrast to conventional $2D$
Josephson arrays, vortices are not independent degrees of freedom in
this network as they are linearly confined.
\bibitem{ferro} Note that for $\alpha \lesssim 1/N$ ferromagnetic
ordering results.
For $\alpha=0$ the system corresponds to the conventional
infinite-range ferromagnet albeit that $J_0$ must itself acquire an
extra $1/\sqrt{N}$ dependence to yield the usual $N$-scaling of
physical observables.
\bibitem{numerics}We checked numerically that these results are
correct for finite-size systems ($N \sim 50$) as well.
\bibitem{disclaimer} We have not explicitly calculated all the states
beneath the transition but it seems highly probable that the
degeneracy of the soft mode will lead to a corresponding degeneracy of
states.
\bibitem{details} To derive the discontinuity in $|m_i|^2$ we evaluate
the free energy $F\{m_i\}$ to all orders in $q=\frac{1}{2N} \sum
|m_i|^2$ but expand in $(|m_i|^2-q)$. Finally, $q$ itself is obtained
by minimizing $F(q)$.
\end{references}
\end{document}